\documentclass[sigconf]{acmart}

\AtBeginDocument{%
  \providecommand\BibTeX{{%
    \normalfont B\kern-0.5em{\scshape i\kern-0.25em b}\kern-0.8em\TeX}}}

\acmConference[ImpactRS '19]{ImpactRS '19: 1st Workshop on the Impact of Recommender Systems}
{September 19, 2019}{Copenhagen, Denmark\\
Copyright \textcopyright 2019 for this paper by its authors. Use permitted under Creative Commons License Attribution 4.0 International (CC BY 4.0).\\ \includegraphics[scale=0.8]{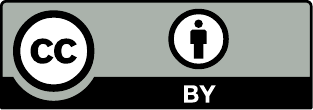}} 
\acmPrice{}
\acmISBN{}
\acmYear{}

\setcopyright{none}
\makeatletter
\renewcommand\@formatdoi[1]{\ignorespaces}
\makeatother

\usepackage{fancyhdr}

\newif\ifworkinprogress
\workinprogresstrue

\ifworkinprogress
	\newcommand{\ez}[1]{\textcolor{green}{\textbf{[Eva] #1}}}
	\newcommand{\chb}[1]{\textcolor{magenta}{\textbf{[Christine] #1}}}
\else
  \newcommand{\ez}[1]{}
  \newcommand{\chb}[1]{}
\fi

\begin{document}

\title[Leveraging Multi-Method Evaluation for Multi-Stakeholder Settings]{Leveraging Multi-Method Evaluation for\\Multi-Stakeholder Settings}
\author{Christine Bauer}
\affiliation{
  \institution{Johannes Kepler University Linz, Austria}
}
\email{christine.bauer@jku.at}

\author{Eva Zangerle}
\affiliation{
  \institution{University of Innsbruck, Austria}
  }
\email{eva.zangerle@uibk.ac.at}

\renewcommand{\shortauthors}{Bauer and Zangerle}

\begin{abstract}
  In this paper, we focus on recommendation settings with multiple stakeholders with possibly varying goals and interests, and argue that a single evaluation method or measure is not able to evaluate all relevant aspects 
  in such a complex setting. We 
  reason that employing a multi-method evaluation, where multiple evaluation methods or measures are combined and integrated, allows for getting a richer picture and 
  prevents
  blind spots in the evaluation outcome.
\end{abstract}



\maketitle

\fancyhead{}
\pagestyle{fancy}
\fancyhf{}
\lhead{ImpactRS '19, September 19, 2019, Copenhagen, Denmark}
\rhead{Bauer and Zangerle}
\rfoot{\thepage}

\section{Introduction}

In recommender systems (RS) research, we observe a strong focus on advancing systems such that they accurately predict items that an individual user may be interested in. The approach of evaluating an RS is thereby largely focused on system-centric methods and metrics (e.g., recall and precision 
in 
leave-$n$-out analyses~\cite{knijnenburg2015evaluating}). 
By employing such an evaluation approach and 
aiming at optimizing these metrics, 
the following crucial components in the ecosystem 
are neglected~\cite{burke2017multisided,Ekstrand2016_behaviorism}: 
(i)~multiple stakeholders are embedded in the ecosystem, but current research largely considers merely the role of the end consumer; 
(ii)~the stakeholders typically have diverging interests and objectives for an RS; however, accurately predicting a user's interests is the predominant focus in current RS research; and 
(iii)~with taking a mainly accuracy-driven, system-centric approach to evaluation, many aspects that determine a user's experience with an RS 
are not considered~\cite{knijnenburg2015evaluating}. 
This results in an incomplete picture of user experience, leaving ``blind spots'' that are not captured in the quality evaluation of an RS. 
Although studies~\cite{Azaria:2013:MRS:2507157.2507162} could show that 
a lower accuracy rate may increase the business utility (e.g., revenue) without any significant drop in user satisfaction, the objectives and interests of stakeholders other than the user are typically not the focus of research in academic settings in the RS community.


In this 
paper, we call for considering the multiple stakeholders in RS evaluation and postulate that only taking a multi-method evaluation approach allows for capturing and assessing the various interests, objectives, and experiences of these very stakeholders; thus, contributing to eliminating the blind spots in RS evaluation.

To illustrate the opportunities of multi-method evaluations in evaluating an RS from multiple stakeholders' perspectives, we exemplarily focus on the music domain. 
In Section~\ref{sec:related}, we outline related work.
Section~\ref{sec:stakeholders} points out the multitude and diversity of stakeholders 
to be considered in the evaluation of music RS. 
In Section~\ref{sec:evaluations}, we exemplify for selected stakeholders
how the integration of multiple evaluation methods contributes to eliminating blind spots in RS evaluation while balancing the stakeholders' interests.


\section{Related Work}\label{sec:related}

The idea to combine different research methods is not a new one.
The concept of mixed methods research~\cite{creswell2003}, for instance, combines quantitative and qualitative research approaches. It has been termed the third methodological paradigm, with quantitative and qualitative methods representing the first and second paradigm
respectively~\cite{teddlie2009foundations}. 
Yet, 
it seems that mixed methods research appears to attract considerable interest but is rarely brought into practice~\cite{aagerfalk2013embracing}. 
From a practical point of view, the reasons for the low adoption of evaluations leveraging multiple methods are manifold, including higher costs, higher complexity, wider 
skill requirements compared to adopting one method only~\cite{Celik:2018:UIU:3213586.3226202}.

For RS research, \citet{gunawardana2015evaluating} point out that there is an extensive number of aspects 
that may be considered when assessing the performance of a recommendation algorithm.
Indeed, already early research on RS pointed towards the wide variety of metrics available for system-centric RS evaluation, including classification metrics, predictive metrics, coverage metrics, confidence metrics, and learning rate metrics~\cite{said2012recommender}. As accuracy-driven evaluation has been shown not to be able to 
capture all the aspects that are relevant for  
user satisfaction~\cite{konstan2012recommender}, more user-relevant metrics and measures have been introduced and considered over time~\cite{kaminskas2016diversity} (so-called ``quality factors beyond accuracy''~\cite{Jannach:2016:RSB:3013530.2891406}). 
This wider range of objectives 
includes qualities such as 
novelty, serendipity~\cite{Herlocker:2004:ECF:963770.963772}, or 
diversity~\cite{kaminskas2016diversity}. 

\citet{kohavi2013online} stress the importance of applying multiple metrics also in the field of A/B testing and online experiments, pointing out that different metrics reflect different concerns.
For A/B testing in RS research, \citet{Ekstrand2016_behaviorism} emphasize the need to include methods and metrics that go beyond the typical A/B behavior metrics. They argue that the currently dominating RS evaluation based on implicit feedback and A/B testing (they refer to this combination as ``behaviorism'') is often very limited in its ability to explain why users acted in a particular way. They emphasize that experiments need to be thoroughly grounded in theory and point to the advantages of collecting subjective responses from users which may help to explain their behavior.


\citet{jannach2017price} point out that academic research in the field of RS tends to focus on the consumer's perspective with the goal to maximize the consumer's utility (measured in terms of the most accurate items for a user), while maximizing the provider's utility (e.g., in terms of profit) appears to be neglected.
While industry research on RS will naturally build around the provider's perspective, publications in this area are scarce~\cite{ZANKER2019_pastpresentfuture}.

\section{Digital Music Stakeholders} 
\label{sec:stakeholders}
Various stakeholders are involved in the digital music value chain~\cite{abdollahpouri2017multiple}. 
From songwriters 
who create songs; 
to performers (e.g., (solo) artists, bands); 
to music producers who take a broad role in the 
production of a track; to record companies, including the three major labels; 
to music platform providers with huge repositories of music tracks, 
acting at the interface to music consumers; and hundreds of millions of music consumers 
with different music preferences and various objectives for using 
RS (e.g., discovering previously 
unknown items, rediscovering items not having listened to in a while 
); to society at large with its social, 
economic, and political 
objectives and needs.



Some stakeholders 
focus on user experience, where the 
goal is to propose ``the right music, to the right user, at the right moment''~\cite{laplante2014_improving}.
Other stakeholders have business-oriented utility functions
~\cite{abdollahpouri2017multiple}. For instance, artists 
will most likely want to have their own songs recommended to consumers. While some artists may be fine with any of their songs being recommended, others 
may prefer to increase the playcount of a particular song 
(e.g., 
to reach the top charts, which would open an opportunity to draw 
an even broader audience; 
or 
some song may generate higher revenues than others due to contract rules). Achieving additional 1,000 playcounts will not get apparent for highly popular 
artists 
with 
yearly playcounts of several billions, 
but could be an important milestone for 
a comparatively 
less popular (local) artist.

\section{Balancing Stakeholder Interests in Evaluation}\label{sec:evaluations}
In the following, we aim to make the case for multi-method evaluations that contribute to identifying the strong and weak spots of a music RS for the stakeholders involved, where we focus on the users' and artists' perspectives in this section.



From a user perspective, recommendations that are adequate in terms of system-centric measures---e.g., the predictive accuracy of recommendation algorithms---do not necessarily meet a user's expectations~\cite{konstan2012recommender}. 
User-centric evaluation methods, in contrast, involve users who interact with an RS~\cite{knijnenburg2015evaluating} to gather user feedback
~\cite{Beel:2013:online_offline} 
either implicitly or explicitly (depending on the concrete evaluation design).
Such methods measure a user's perceived quality of the RS at the time of recommendation, e.g., by established questionnaires~\cite{pu2011user}. Still, relying solely on user-centric methods does not reveal the accuracy of the recommendations, because, given the vast amount of items, 
users are not able to judge 
whether a given recommendation was indeed the most relevant one~\cite{Beel:2013:online_offline}. 

Measuring accuracy does not capture the recommendations' usefulness for users, because higher accuracy scores do not necessarily imply higher user satisfaction~\cite{McNee:2006:AEA:1125451.1125659}. 
For instance, a user's most favorite song is an accurate prediction; still, repeating the same song five times is, though accurate, likely not a satisfying experience. Hence, we argue that for evaluating the user's perspective of a RS---the user being only one of the many stakeholders involved---multiple evaluation methods and measures are required. This may include 
combining a set of different measures (ranging from recall and precision to serendipity, list diversity or novelty) or integrating different evaluation methods (ranging from leave-n-out offline experiments to user studies and A/B testing). Furthermore, although A/B-testing using user's implicit feedback is 
effective for testing the impact of different algorithms or designs on user behavior---and is, thus, frequently considered the ``golden standard'' for recommender evaluation---, it has limited 
ability in explaining why users acted in a particular way~\cite{Ekstrand2016_behaviorism}. Additional information (e.g., users' subjective responses) is necessary to allow for explaining behavior.

In short, sticking to a single evaluation method narrows our view on the RS, literally having blinders on, while devising and evaluating RS. We can borrow from 
social and behavioral sciences, 
where, e.g., mixed-methods research combines quantitative and qualitative evaluations 
using different designs~\cite{creswell2003}. Creswell's proposed designs include---among others---the convergent parallel design and the sequential design. In the convergent parallel design, two evaluation methods are first applied in parallel, and finally integrated into a single interpretation. 
The sequential design uses sequential timing, employing the methods in distinct phases. The second phase of the study, using the second method, is designed such that it follows from the results of the first phase. Depending on the research goal and the concrete choice of methods, researchers may either interpret how the second phase's results help to explain the initial results (explanatory design) or they build on the exploratory results of the first phase to subsequently employ a different method (in the second phase) to test or generalize the initial findings (exploratory design). For instance,~\citet{Kamehkhosh:2017:UPN:3079628.3079668} showed that in the field of music RS, the results of a conducted offline evaluation could be reproduced with online studies assessing the users' perceived quality of recommendations. Similarly, for the Recommender Systems Challenge 2017, participants firstly evaluated their prototypes in offline evaluations, before actually deploying them and evaluating them in the live system utilizing A/B tests~\cite{abel2017recsys}, showing that many 
of RS who performed well in offline evaluations were able to repeat this in online experiments. However, some of the devised RS also performed substantially worse in online experiments---highlighting a shortcoming that was not revealed by evaluating from solely an offline perspective. Along the same lines, Ekstrand and Willemsen~\cite{Ekstrand2016_behaviorism} state that utilizing behaviorism for evaluation purposes (e.g., through A/B tests) is not sufficient 
to understand \emph{why} users act 
in a particular way and, for instance, 
like a particular recommendation.

While academic research in the field of RS tends to focus on maximizing the users' utility, some authors (e.g.,~\cite{jannach2017price}) emphasize the importance of 
profit (or value) maximization. 
Profit 
maximization may not only be a goal for platform providers, but also for artists who are the content providers for music platforms.
From an artist's perspective, a good RS recommends the respective artist's songs sufficiently frequently, which may ultimately lead to playcounts, likes, purchases, profit maximization, etc. Evaluating for profit may, though, leave blind spots. For example, depending on the 
chosen strategy, 
an artist may want to emphasize other values such as expanding the audience (thus, reaching new listeners) or increasing the listening or purchase volume within the current fan base. Hence, metrics such as 
number of unique listeners per artist, 
the sum of playcounts over all songs of an artist, 
and metrics such as profit-per-audience type may be valuable for RS optimization and need to be considered in the RS evaluation strategy. Accordingly, evaluation efforts need to elicit and integrate the artists' goals and preferences  need to be elicited and integrated into the evaluation efforts. 
While evaluation on a per-artist-basis might be interesting for the individual artists (e.g., for a comparison between platforms and their integrated RS), 
it may not be adequate for an overall RS evaluation. Still, an RS needs to be evaluated for its ability to serve the various strategies and for revealing potential tendencies towards the one or other strategy. As the targeted strategy might correlate with artist characteristics (e.g., top-of-the-top artists vs. ``long tail'' artist; early career vs. come-back phase vs. long-term career; mainstream artists vs. 
niche genres), it might be in the society's interest to evaluate for and ensure a balance.

Having given these examples, we 
emphasize that, due to interdependencies between the RS and the various stakeholders' actions, 
the entire RS ecosystem has to be taken into account in the evaluation.
For instance, low accuracy of recommendations and low user experience are not likely to continuously increase profits for the platform provider and all kinds of artists; high accuracy does not automatically imply high user experience and may not contribute to profit maximization.

\section{Conclusions}
In this position paper, we exemplarily focused on the digital music ecosystem to illustrate that multiple stakeholders are impacted by music RS, and discussed the opportunities of multi-method evaluations to consider the multiple stakeholders' perspectives.
We emphasize that---irrespective of the application domain---there are always multiple stakeholders involved in recommendation settings. Hence, there are always multiple---and possibly diverging---perspectives and goals of these very stakeholders which need to be considered in evaluating an RS. 
Consequently, multiple evaluation methods and criteria have to be combined and possibly also weighted. 

\emph{Multi-method evaluations} allow for gathering a richer and more integrated picture of the quality of a RS and contributes to understanding the various phenomena involved in a multi-stakeholder setting, for which one method in isolation would be insufficient~\cite{venkatesh2013bridging}. 

\begin{acks}
This research is supported by the Austrian Science Fund (FWF): V579.
\end{acks}

\balance
\bibliographystyle{ACM-Reference-Format}
\bibliography{references}


\begin{thebibliography}{25}


\ifx \showCODEN    \undefined \def \showCODEN     #1{\unskip}     \fi
\ifx \showDOI      \undefined \def \showDOI       #1{#1}\fi
\ifx \showISBNx    \undefined \def \showISBNx     #1{\unskip}     \fi
\ifx \showISBNxiii \undefined \def \showISBNxiii  #1{\unskip}     \fi
\ifx \showISSN     \undefined \def \showISSN      #1{\unskip}     \fi
\ifx \showLCCN     \undefined \def \showLCCN      #1{\unskip}     \fi
\ifx \shownote     \undefined \def \shownote      #1{#1}          \fi
\ifx \showarticletitle \undefined \def \showarticletitle #1{#1}   \fi
\ifx \showURL      \undefined \def \showURL       {\relax}        \fi
\providecommand\bibfield[2]{#2}
\providecommand\bibinfo[2]{#2}
\providecommand\natexlab[1]{#1}
\providecommand\showeprint[2][]{arXiv:#2}

\bibitem[\protect\citeauthoryear{Abdollahpouri and Essinger}{Abdollahpouri and
  Essinger}{2017}]%
        {abdollahpouri2017multiple}
\bibfield{author}{\bibinfo{person}{Himan Abdollahpouri} {and}
  \bibinfo{person}{Steve Essinger}.} \bibinfo{year}{2017}\natexlab{}.
\newblock \showarticletitle{Multiple stakeholders in music recommender
  systems}. In \bibinfo{booktitle}{\emph{1st International Workshop on
  Value-Aware and Multistakeholder Recommendation at RecSys 2017}}
  \emph{(\bibinfo{series}{VAMS '17})}.
\newblock
\showeprint[arxiv]{1708.00120}


\bibitem[\protect\citeauthoryear{Abel, Deldjoo, Elahi, and Kohlsdorf}{Abel
  et~al\mbox{.}}{2017}]%
        {abel2017recsys}
\bibfield{author}{\bibinfo{person}{Fabian Abel}, \bibinfo{person}{Yashar
  Deldjoo}, \bibinfo{person}{Mehdi Elahi}, {and} \bibinfo{person}{Daniel
  Kohlsdorf}.} \bibinfo{year}{2017}\natexlab{}.
\newblock \showarticletitle{Recsys challenge 2017: Offline and online
  evaluation}. In \bibinfo{booktitle}{\emph{Proceedings of the Eleventh ACM
  Conference on Recommender Systems}}. \bibinfo{publisher}{ACM},
  \bibinfo{address}{New York, NY, USA}, \bibinfo{pages}{372--373}.
\newblock


\bibitem[\protect\citeauthoryear{{\AA}gerfalk}{{\AA}gerfalk}{2013}]%
        {aagerfalk2013embracing}
\bibfield{author}{\bibinfo{person}{P{\"a}r~J {\AA}gerfalk}.}
  \bibinfo{year}{2013}\natexlab{}.
\newblock \showarticletitle{Embracing diversity through mixed methods
  research}.
\newblock \bibinfo{journal}{\emph{European Journal of Information Systems}}
  \bibinfo{volume}{22}, \bibinfo{number}{3} (\bibinfo{year}{2013}),
  \bibinfo{pages}{251--256}.
\newblock
\showISSN{1476-9344}
\urldef\tempurl%
\url{https://doi.org/10.1057/ejis.2013.6}
\showDOI{\tempurl}


\bibitem[\protect\citeauthoryear{Azaria, Hassidim, Kraus, Eshkol, Weintraub,
  and Netanely}{Azaria et~al\mbox{.}}{2013}]%
        {Azaria:2013:MRS:2507157.2507162}
\bibfield{author}{\bibinfo{person}{Amos Azaria}, \bibinfo{person}{Avinatan
  Hassidim}, \bibinfo{person}{Sarit Kraus}, \bibinfo{person}{Adi Eshkol},
  \bibinfo{person}{Ofer Weintraub}, {and} \bibinfo{person}{Irit Netanely}.}
  \bibinfo{year}{2013}\natexlab{}.
\newblock \showarticletitle{Movie Recommender System for Profit Maximization}.
  In \bibinfo{booktitle}{\emph{Proceedings of the 7th ACM Conference on
  Recommender Systems}} \emph{(\bibinfo{series}{RecSys '13})}.
  \bibinfo{publisher}{ACM}, \bibinfo{address}{New York, NY, USA},
  \bibinfo{pages}{121--128}.
\newblock
\showISBNx{978-1-4503-2409-0}
\urldef\tempurl%
\url{https://doi.org/10.1145/2507157.2507162}
\showDOI{\tempurl}


\bibitem[\protect\citeauthoryear{Beel, Genzmehr, Langer, N\"{u}rnberger, and
  Gipp}{Beel et~al\mbox{.}}{2013}]%
        {Beel:2013:online_offline}
\bibfield{author}{\bibinfo{person}{Joeran Beel}, \bibinfo{person}{Marcel
  Genzmehr}, \bibinfo{person}{Stefan Langer}, \bibinfo{person}{Andreas
  N\"{u}rnberger}, {and} \bibinfo{person}{Bela Gipp}.}
  \bibinfo{year}{2013}\natexlab{}.
\newblock \showarticletitle{A Comparative Analysis of Offline and Online
  Evaluations and Discussion of Research Paper Recommender System Evaluation}.
  In \bibinfo{booktitle}{\emph{Proceedings of the International Workshop on
  Reproducibility and Replication in Recommender Systems Evaluation}}
  \emph{(\bibinfo{series}{RepSys '13})}. \bibinfo{publisher}{ACM},
  \bibinfo{address}{New York, NY, USA}, \bibinfo{pages}{7--14}.
\newblock
\showISBNx{978-1-4503-2465-6}
\urldef\tempurl%
\url{https://doi.org/10.1145/2532508.2532511}
\showDOI{\tempurl}


\bibitem[\protect\citeauthoryear{Burke}{Burke}{2017}]%
        {burke2017multisided}
\bibfield{author}{\bibinfo{person}{Robin Burke}.}
  \bibinfo{year}{2017}\natexlab{}.
\newblock \showarticletitle{Multisided fairness for recommendation}. In
  \bibinfo{booktitle}{\emph{4th Workshop on Fairness, Accountability, and
  Transparency in Machine Learning}} \emph{(\bibinfo{series}{FAT/ML '17})}.
\newblock
\showeprint[arxiv]{1707.00093}


\bibitem[\protect\citeauthoryear{Celik, Torre, Koceva, Bauer, Zangerle, and
  Knijnenburg}{Celik et~al\mbox{.}}{2018}]%
        {Celik:2018:UIU:3213586.3226202}
\bibfield{author}{\bibinfo{person}{Ilknur Celik}, \bibinfo{person}{Ilaria
  Torre}, \bibinfo{person}{Frosina Koceva}, \bibinfo{person}{Christine Bauer},
  \bibinfo{person}{Eva Zangerle}, {and} \bibinfo{person}{Bart Knijnenburg}.}
  \bibinfo{year}{2018}\natexlab{}.
\newblock \showarticletitle{UMAP 2018 Intelligent User-Adapted Interfaces:
  Design and Multi-Modal Evaluation (IUadaptMe)}. In
  \bibinfo{booktitle}{\emph{Adjunct Publication of the 26th Conference on User
  Modeling, Adaptation and Personalization}} \emph{(\bibinfo{series}{UMAP
  '18})}. \bibinfo{publisher}{ACM}, \bibinfo{address}{New York, NY, USA},
  \bibinfo{pages}{137--139}.
\newblock
\showISBNx{978-1-4503-5784-5}
\urldef\tempurl%
\url{https://doi.org/10.1145/3213586.3226202}
\showDOI{\tempurl}


\bibitem[\protect\citeauthoryear{Creswell}{Creswell}{2003}]%
        {creswell2003}
\bibfield{author}{\bibinfo{person}{John~W. Creswell}.}
  \bibinfo{year}{2003}\natexlab{}.
\newblock \bibinfo{booktitle}{\emph{Research design: qualitative, quantitative,
  and mixed methods approaches} (\bibinfo{edition}{2nd} ed.)}.
\newblock \bibinfo{publisher}{Sage Publications}, \bibinfo{address}{Thousand
  Oaks, CA, USA}.
\newblock


\bibitem[\protect\citeauthoryear{Ekstrand and Willemsen}{Ekstrand and
  Willemsen}{2016}]%
        {Ekstrand2016_behaviorism}
\bibfield{author}{\bibinfo{person}{Michael~D. Ekstrand} {and}
  \bibinfo{person}{Martijn~C. Willemsen}.} \bibinfo{year}{2016}\natexlab{}.
\newblock \showarticletitle{Behaviorism is not enough: better recommendations
  through listening to users}. In \bibinfo{booktitle}{\emph{Proceedings of the
  10th ACM Conference on Recommender Systems}} \emph{(\bibinfo{series}{RecSys
  '16})}. \bibinfo{publisher}{ACM}, \bibinfo{address}{New York, NY, USA},
  \bibinfo{pages}{221--224}.
\newblock
\showISBNx{978-1-4503-4035-9}
\urldef\tempurl%
\url{https://doi.org/10.1145/2959100.2959179}
\showDOI{\tempurl}


\bibitem[\protect\citeauthoryear{Gunawardana and Shani}{Gunawardana and
  Shani}{2015}]%
        {gunawardana2015evaluating}
\bibfield{author}{\bibinfo{person}{Asela Gunawardana} {and}
  \bibinfo{person}{Guy Shani}.} \bibinfo{year}{2015}\natexlab{}.
\newblock \showarticletitle{Evaluating Recommender Systems}.
\newblock In \bibinfo{booktitle}{\emph{Recommender Systems Handbook}
  (\bibinfo{edition}{2nd} ed.)}, \bibfield{editor}{\bibinfo{person}{Francesco
  Ricci}, \bibinfo{person}{Lior Rokach}, {and} \bibinfo{person}{Bracha
  Shapira}} (Eds.). \bibinfo{publisher}{Springer}, \bibinfo{address}{Boston,
  MA, USA}, \bibinfo{pages}{265--308}.
\newblock
\showISBNx{978-1-4899-7637-6}
\urldef\tempurl%
\url{https://doi.org/10.1007/978-1-4899-7637-6_8}
\showDOI{\tempurl}


\bibitem[\protect\citeauthoryear{Herlocker, Konstan, Terveen, and
  Riedl}{Herlocker et~al\mbox{.}}{2004}]%
        {Herlocker:2004:ECF:963770.963772}
\bibfield{author}{\bibinfo{person}{Jonathan~L. Herlocker},
  \bibinfo{person}{Joseph~A. Konstan}, \bibinfo{person}{Loren~G. Terveen},
  {and} \bibinfo{person}{John~T. Riedl}.} \bibinfo{year}{2004}\natexlab{}.
\newblock \showarticletitle{Evaluating Collaborative Filtering Recommender
  Systems}.
\newblock \bibinfo{journal}{\emph{ACM Transaction on Information Systems}}
  \bibinfo{volume}{22}, \bibinfo{number}{1} (\bibinfo{date}{Jan.}
  \bibinfo{year}{2004}), \bibinfo{pages}{5--53}.
\newblock
\showISSN{1046-8188}
\urldef\tempurl%
\url{https://doi.org/10.1145/963770.963772}
\showDOI{\tempurl}


\bibitem[\protect\citeauthoryear{Jannach and Adomavicius}{Jannach and
  Adomavicius}{2017}]%
        {jannach2017price}
\bibfield{author}{\bibinfo{person}{Dietmar Jannach} {and}
  \bibinfo{person}{Gediminas Adomavicius}.} \bibinfo{year}{2017}\natexlab{}.
\newblock \showarticletitle{Price and profit awareness in recommender systems}.
  In \bibinfo{booktitle}{\emph{1st International Workshop on Value-Aware and
  Multistakeholder Recommendation at RecSys 2017}} \emph{(\bibinfo{series}{VAMS
  '17})}.
\newblock
\showeprint[arxiv]{1707.08029}


\bibitem[\protect\citeauthoryear{Jannach, Resnick, Tuzhilin, and
  Zanker}{Jannach et~al\mbox{.}}{2016}]%
        {Jannach:2016:RSB:3013530.2891406}
\bibfield{author}{\bibinfo{person}{Dietmar Jannach}, \bibinfo{person}{Paul
  Resnick}, \bibinfo{person}{Alexander Tuzhilin}, {and} \bibinfo{person}{Markus
  Zanker}.} \bibinfo{year}{2016}\natexlab{}.
\newblock \showarticletitle{Recommender Systems --- Beyond Matrix Completion}.
\newblock \bibinfo{journal}{\emph{Commun. ACM}} \bibinfo{volume}{59},
  \bibinfo{number}{11} (\bibinfo{year}{2016}), \bibinfo{pages}{94--102}.
\newblock
\showISSN{0001-0782}
\urldef\tempurl%
\url{https://doi.org/10.1145/2891406}
\showDOI{\tempurl}


\bibitem[\protect\citeauthoryear{Kamehkhosh and Jannach}{Kamehkhosh and
  Jannach}{2017}]%
        {Kamehkhosh:2017:UPN:3079628.3079668}
\bibfield{author}{\bibinfo{person}{Iman Kamehkhosh} {and}
  \bibinfo{person}{Dietmar Jannach}.} \bibinfo{year}{2017}\natexlab{}.
\newblock \showarticletitle{User Perception of Next-Track Music
  Recommendations}. In \bibinfo{booktitle}{\emph{Proceedings of the 25th
  Conference on User Modeling, Adaptation and Personalization}}
  \emph{(\bibinfo{series}{UMAP '17})}. \bibinfo{publisher}{ACM},
  \bibinfo{address}{New York, NY, USA}, \bibinfo{pages}{113--121}.
\newblock
\showISBNx{978-1-4503-4635-1}
\urldef\tempurl%
\url{https://doi.org/10.1145/3079628.3079668}
\showDOI{\tempurl}


\bibitem[\protect\citeauthoryear{Kaminskas and Bridge}{Kaminskas and
  Bridge}{2016}]%
        {kaminskas2016diversity}
\bibfield{author}{\bibinfo{person}{Marius Kaminskas} {and}
  \bibinfo{person}{Derek Bridge}.} \bibinfo{year}{2016}\natexlab{}.
\newblock \showarticletitle{Diversity, Serendipity, Novelty, and Coverage: A
  Survey and Empirical Analysis of Beyond-Accuracy Objectives in Recommender
  Systems}.
\newblock \bibinfo{journal}{\emph{ACM Transactions on Interactive Intelligent
  Systems}} \bibinfo{volume}{7}, \bibinfo{number}{1}, Article
  \bibinfo{articleno}{2} (\bibinfo{year}{2016}), \bibinfo{numpages}{42}~pages.
\newblock
\showISSN{2160-6455}
\urldef\tempurl%
\url{https://doi.org/10.1145/2926720}
\showDOI{\tempurl}


\bibitem[\protect\citeauthoryear{Knijnenburg and Willemsen}{Knijnenburg and
  Willemsen}{2015}]%
        {knijnenburg2015evaluating}
\bibfield{author}{\bibinfo{person}{Bart~P. Knijnenburg} {and}
  \bibinfo{person}{Martijn~C. Willemsen}.} \bibinfo{year}{2015}\natexlab{}.
\newblock \showarticletitle{Evaluating Recommender Systems with User
  Experiments}.
\newblock In \bibinfo{booktitle}{\emph{Recommender Systems Handbook}
  (\bibinfo{edition}{2nd} ed.)}, \bibfield{editor}{\bibinfo{person}{Francesco
  Ricci}, \bibinfo{person}{Lior Rokach}, {and} \bibinfo{person}{Bracha
  Shapira}} (Eds.). \bibinfo{publisher}{Springer}, \bibinfo{address}{Boston,
  MA, USA}, \bibinfo{pages}{309--352}.
\newblock
\showISBNx{978-1-4899-7637-6}
\urldef\tempurl%
\url{https://doi.org/10.1007/978-1-4899-7637-6_9}
\showDOI{\tempurl}


\bibitem[\protect\citeauthoryear{Kohavi, Deng, Frasca, Walker, Xu, and
  Pohlmann}{Kohavi et~al\mbox{.}}{2013}]%
        {kohavi2013online}
\bibfield{author}{\bibinfo{person}{Ron Kohavi}, \bibinfo{person}{Alex Deng},
  \bibinfo{person}{Brian Frasca}, \bibinfo{person}{Toby Walker},
  \bibinfo{person}{Ya Xu}, {and} \bibinfo{person}{Nils Pohlmann}.}
  \bibinfo{year}{2013}\natexlab{}.
\newblock \showarticletitle{Online Controlled Experiments at Large Scale}. In
  \bibinfo{booktitle}{\emph{Proceedings of the 19th ACM SIGKDD International
  Conference on Knowledge Discovery and Data Mining}}
  \emph{(\bibinfo{series}{KDD '13})}. \bibinfo{publisher}{ACM},
  \bibinfo{address}{New York, NY, USA}, \bibinfo{pages}{1168--1176}.
\newblock
\showISBNx{978-1-4503-2174-7}
\urldef\tempurl%
\url{https://doi.org/10.1145/2487575.2488217}
\showDOI{\tempurl}


\bibitem[\protect\citeauthoryear{Konstan and Riedl}{Konstan and Riedl}{2012}]%
        {konstan2012recommender}
\bibfield{author}{\bibinfo{person}{Joseph~A. Konstan} {and}
  \bibinfo{person}{John Riedl}.} \bibinfo{year}{2012}\natexlab{}.
\newblock \showarticletitle{Recommender systems: from algorithms to user
  experience}.
\newblock \bibinfo{journal}{\emph{User Modeling and User-Adapted Interaction}}
  \bibinfo{volume}{22}, \bibinfo{number}{1} (\bibinfo{year}{2012}),
  \bibinfo{pages}{101--123}.
\newblock
\showISSN{1573-1391}
\urldef\tempurl%
\url{https://doi.org/10.1007/s11257-011-9112-x}
\showDOI{\tempurl}


\bibitem[\protect\citeauthoryear{Laplante}{Laplante}{2014}]%
        {laplante2014_improving}
\bibfield{author}{\bibinfo{person}{Audrey Laplante}.}
  \bibinfo{year}{2014}\natexlab{}.
\newblock \showarticletitle{Improving music recommender systems: what can we
  learn from research on music tags?}. In \bibinfo{booktitle}{\emph{15th
  International Society for Music Information Retrieval Conference}}
  \emph{(\bibinfo{series}{ISMIR '14})}. International Society for Music
  Information Retrieval, \bibinfo{pages}{451--456}.
\newblock


\bibitem[\protect\citeauthoryear{McNee, Riedl, and Konstan}{McNee
  et~al\mbox{.}}{2006}]%
        {McNee:2006:AEA:1125451.1125659}
\bibfield{author}{\bibinfo{person}{Sean~M. McNee}, \bibinfo{person}{John
  Riedl}, {and} \bibinfo{person}{Joseph~A. Konstan}.}
  \bibinfo{year}{2006}\natexlab{}.
\newblock \showarticletitle{Being Accurate is Not Enough: How Accuracy Metrics
  Have Hurt Recommender Systems}. In \bibinfo{booktitle}{\emph{CHI '06 Extended
  Abstracts on Human Factors in Computing Systems}} \emph{(\bibinfo{series}{CHI
  EA '06})}. \bibinfo{publisher}{ACM}, \bibinfo{address}{New York, NY, USA},
  \bibinfo{pages}{1097--1101}.
\newblock
\showISBNx{1-59593-298-4}
\urldef\tempurl%
\url{https://doi.org/10.1145/1125451.1125659}
\showDOI{\tempurl}


\bibitem[\protect\citeauthoryear{Pu, Chen, and Hu}{Pu et~al\mbox{.}}{2011}]%
        {pu2011user}
\bibfield{author}{\bibinfo{person}{Pearl Pu}, \bibinfo{person}{Li Chen}, {and}
  \bibinfo{person}{Rong Hu}.} \bibinfo{year}{2011}\natexlab{}.
\newblock \showarticletitle{A User-centric Evaluation Framework for Recommender
  Systems}. In \bibinfo{booktitle}{\emph{Proceedings of the 5th ACM Conference
  on Recommender Systems}} \emph{(\bibinfo{series}{RecSys '11})}.
  \bibinfo{publisher}{ACM}, \bibinfo{address}{New York, NY, USA},
  \bibinfo{pages}{157--164}.
\newblock
\showISBNx{978-1-4503-0683-6}
\urldef\tempurl%
\url{https://doi.org/10.1145/2043932.2043962}
\showDOI{\tempurl}


\bibitem[\protect\citeauthoryear{Said, Tikk, Stumpf, Shi, Larson, and
  Cremonesi}{Said et~al\mbox{.}}{2012}]%
        {said2012recommender}
\bibfield{author}{\bibinfo{person}{Alan Said}, \bibinfo{person}{Domonkos Tikk},
  \bibinfo{person}{Klara Stumpf}, \bibinfo{person}{Yue Shi},
  \bibinfo{person}{Martha Larson}, {and} \bibinfo{person}{Paolo Cremonesi}.}
  \bibinfo{year}{2012}\natexlab{}.
\newblock \showarticletitle{Recommender Systems Evaluation: A 3D Benchmark}. In
  \bibinfo{booktitle}{\emph{Proceedings of the Workshop on Recommendation
  Utility Evaluation: Beyond RMSE}} \emph{(\bibinfo{series}{RUE '12})},
  Vol.~\bibinfo{volume}{910}. CEUR Workshop Proceedings,
  \bibinfo{pages}{21--23}.
\newblock
\urldef\tempurl%
\url{http://ceur-ws.org/Vol-910/}
\showURL{%
\tempurl}


\bibitem[\protect\citeauthoryear{Teddlie and Tashakkori}{Teddlie and
  Tashakkori}{2009}]%
        {teddlie2009foundations}
\bibfield{author}{\bibinfo{person}{Charles Teddlie} {and}
  \bibinfo{person}{Abbas Tashakkori}.} \bibinfo{year}{2009}\natexlab{}.
\newblock \bibinfo{booktitle}{\emph{Foundations of mixed methods research:
  Integrating quantitative and qualitative approaches in the social and
  behavioral sciences}}.
\newblock \bibinfo{publisher}{Sage Publications}, \bibinfo{address}{Thousand
  Oaks, CA, USA}.
\newblock


\bibitem[\protect\citeauthoryear{Venkatesh, Brown, and Bala}{Venkatesh
  et~al\mbox{.}}{2013}]%
        {venkatesh2013bridging}
\bibfield{author}{\bibinfo{person}{Viswanath Venkatesh},
  \bibinfo{person}{{Susan A.} Brown}, {and} \bibinfo{person}{Hillol Bala}.}
  \bibinfo{year}{2013}\natexlab{}.
\newblock \showarticletitle{Bridging the qualitative-quantitative divide:
  Guidelines for conducting mixed methods research in information systems}.
\newblock \bibinfo{journal}{\emph{MIS Quarterly}} \bibinfo{volume}{37},
  \bibinfo{number}{1} (\bibinfo{year}{2013}), \bibinfo{pages}{21--54}.
\newblock
\showISSN{0276-7783}


\bibitem[\protect\citeauthoryear{Zanker, Rook, and Jannach}{Zanker
  et~al\mbox{.}}{2019}]%
        {ZANKER2019_pastpresentfuture}
\bibfield{author}{\bibinfo{person}{Markus Zanker}, \bibinfo{person}{Laurens
  Rook}, {and} \bibinfo{person}{Dietmar Jannach}.}
  \bibinfo{year}{2019}\natexlab{}.
\newblock \showarticletitle{Measuring the impact of online personalisation:
  Past, present and future}.
\newblock \bibinfo{journal}{\emph{International Journal of Human-Computer
  Studies}} (\bibinfo{year}{2019}).
\newblock
\showISSN{1071-5819}
\urldef\tempurl%
\url{https://doi.org/10.1016/j.ijhcs.2019.06.006}
\showDOI{\tempurl}


\end{thebibliography}


\end{document}